\RequirePackage{fix-cm}
\documentclass[smallextended]{svjour3}       
\smartqed  
\usepackage{graphicx}
\usepackage{float}
\usepackage{amssymb}
\usepackage{xcolor}
\begin{document}
\title{Simulation of Afshar's Double Slit Experiment
}
\author{Bret Gergely         \and
        Herman Batelaan 
}
\institute{B. Gergely \at
              Department of Physics and Astronomy, University of Nebraska-Lincoln, Theodore P. Jorgensen Hall, Lincoln, NE 68588, USA \\
              \email{bgergely@huskers.unl.edu}           
           \and
           H. Batelaan \at
           Department of Physics and Astronomy, University of Nebraska-Lincoln, Theodore P. Jorgensen Hall, Lincoln, NE 68588, USA \\
           \email{hbatelaan@unl.edu}
}
\date{Received: date / Accepted: date}
\maketitle
\begin{abstract}
Shahriar S. Afshar claimed that his 2007 modified version of the double-slit experiment violates complementarity \cite{Afshar2007}.  He makes two modifications to the standard double-slit experiment.  First, he adds a wire grid that is placed in between the slits and the screen at locations of interference minima.  The second modification is to place a converging lens just after the wire grid.  The idea is that the wire grid implies the existence of interference minima (wave-like behavior), while the lens can simultaneously obtain which-way information (particle-like behavior).  More recently, John G. Cramer \cite{Cramer2016} argued that the experiment bolstered the Transactional Interpretation of Quantum mechanics (TIQM). His argument scrutinizes Bohr's complementarity in favor of TIQM.  We analyze this experiment by simulation using the path integral formulation of quantum mechanics \cite{feynman1948} and find that it agrees with the wave-particle duality relation given by Englert, Greenberg and Yasin (E-G-Y) \cite{Englert1996,GreenbergYasin1988}.  We conclude that the use of Afshar's experiment to provide a testbed for quantum mechanical interpretations is limited.
\keywords{Complementarity \and Transactional Interpretation of Quantum mechanics \and Afshar experiment \and Double-slit \and Which-way}
\end{abstract}

\noindent \textbf{Acknowledgments} We gratefully acknowledge support by the National Science Foundation (NSF) under the award number PHY-1912504. \\
\noindent \textbf{Conflicts of interest} The authors declare no conflict of interests. \\
\noindent \textbf{Availability of data and code} Data and code are available upon request.\\

\section{Introduction}
\label{intro}
In the 2007 paper of Afshar et al. \cite{Afshar2007}, the claim is made that the Englert– Greenberger–Yasin \cite{Englert1996,GreenbergYasin1988} duality relation: $D^2+V^2\leq 1$ is violated (\textit{D} is the Distinguishability and \textit{V}, the Visibility). There have been several analyses of the experiment: analytic \cite{qureshi,steuernagel2007,katzner}, simulation \cite{flores2010,reitzner},  as well a follow up experiment \cite{jacques2008NJP}.   The majority of these analyses reject the original authors' claims.  Yet as recent as 2016, the results of this experiment have been used to bolster support for John Cramer's Transactional Interpretation of Quantum Mechanics (TIQM) \cite{Cramer2016}.  This prompted us to reevaluate the claims. \\
\indent In this paper, Afshar's experiment is simulated using the path integral method \cite{feynman1948,Pritchard,ericpaper}.  This choice allows for the calculation of \textit{D} and \textit{V} prescribed by the original definition \cite{Englert1996}.  This has not been done previously, the simulation of Flores \cite{flores2010} used Fraunhofer diffraction to describe the experiment but found the fringe contrast using a spatial probability distribution.  The simulation of Reitzner \cite{reitzner} did not include results for \textit{D} or \textit{V}.  We find no violation of the E-G-Y relation, and thus no support for TIQM is found. \\
\indent We would like to point out that because our analysis of Afshar's experiment computes the distinguishability $D$ in distinction from previous theoretical analysis, it can be experimentally tested. We predict that a repeat of the experiment with an improved resolution will reveal weak diffraction from the wires and that that diffraction is affecting the value of the distinguishability. \\
\indent For a closed bipartite entangled system, the EGY duality relation becomes a triality relation and distinguishability becomes $D^2=C^2 +P^2$, where $C$ is concurrence and $P$ is predictability \cite{Jakob2010} . These relations are relevant to quantum computing as they identify the relationship between quantum resources \cite{Reis2021,Bassol2021}. For open systems, as in real experiments, the validity of the complementarity relations is limited \cite{Fonseca2014,Souza2016,Georgiev2021}.  Within this context, we hope that our analysis of Afshar's attack underlines the importance of duality relationships. \\

\section{SIMULATION}
\label{sec:1}

\begin{figure}[H]
	\begin{center}
		\textbf{Distinguishibility Measurement}\par\medskip
		\resizebox{0.8\hsize}{!}{\includegraphics*{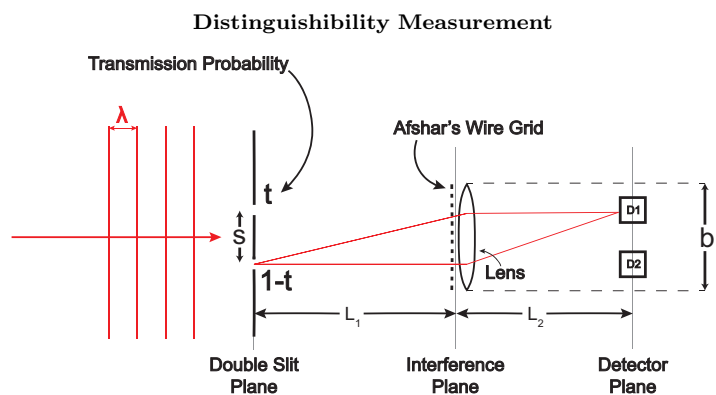}}
		\caption{The setup used by Afshar to determine Distinguishiblity (lens in).  Without Afshar's grid, the detectors can determine to some extent through which slit the particle went, as measured by $D = \frac{D1 - D2}{D1+D2}$. Afshar's idea is that: with a single measurement in the detector plane and a fine grid in the interference plane, \textit{D} will be unaffected, while the interference pattern at the location of the wire grid must have a non-zero contrast.}
		\label{fig:fig1}
	\end{center}
\end{figure}  

\begin{figure}[H]
	\begin{center}
		\textbf{Visibility Measurement}\par\medskip
		\resizebox{0.8\hsize}{!}{\includegraphics*{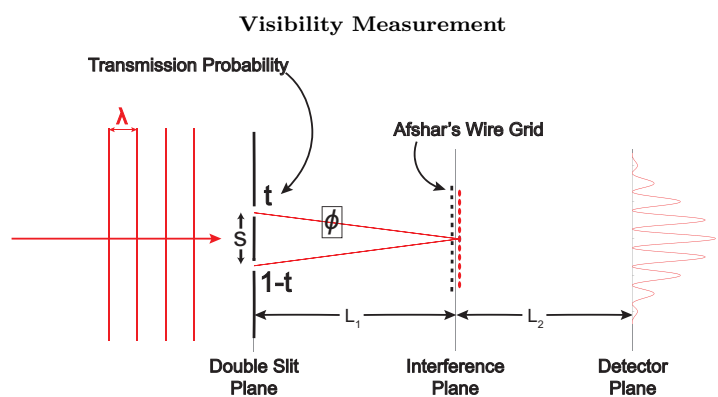}}
		\caption{The setup used by Afshar to determine Visibility (lens out).  The fringe contrast of the interference pattern can be measured in the detector plane (or the interference plane) at any location by varying the phase $\phi$ between the two paths.  This measurement is mutually exclusive with the one indicated in Fig. \ref{fig:fig1}.}
		\label{fig:fig2}
	\end{center}
\end{figure}  
 
\indent The setup for Afshar's experiment is shown in Figs. \ref{fig:fig1} and \ref{fig:fig2}.  A red diode laser (wavelength $\lambda = 650$ $nm$) is incident on two pinholes. Wires with a diameter of 127 $\mu m$ are then placed 0.55 $m$ from the pinholes at locations where diffraction minima would occur if measured (these locations are determined in a separate experiment).  The Distinguishbility measurement (Fig. \ref{fig:fig1}) is performed with a 4x magnification lens placed after the wires.  Detectors D1 and D2 resolve an image of the slits.  Without Afshar's grid, the detectors can determine to some extent, through which slit the particle went. This is quantified by the Distinguishiblty, \textit{D} \cite{Englert1996,jacques2008NJP,jacques2008PRL} \footnote[1]{Note that in this example with point like sources, a simplified definition of D is used (eqn. (\ref{eq:1})).  When inserting the wires, the complete form is needed and is used for our calculations: $D = |P_{1,1}-P_{2,1}|+|P_{1,2}-P_{2,2}|$ where $P_{i,j}$ is the probability to detect at detector \textit{i} following path \textit{j} \cite{jacques2008NJP}.}:

\begin{equation} \label{eq:1}
D = \frac{D_1 - D_2}{D_1+D_2}
\end{equation}

\noindent $D_1$ and $D_2$ are the probabilities to detect at detectors D1 and D2 respectively. \\
\indent The setup for the Visibility measurement is shown in Fig. \ref{fig:fig2}. Here the lens is removed and the intensity is measured.  The Visiblity, \textit{V}, is determined in the following way: 

\begin{equation} \label{eq:2}
V = \frac{I_{max} - I_{min}}{I_{max}+I_{min}}
\end{equation}

\noindent $I_{max}$ and $I_{min}$ can be determined at any location in the detector plane by varying the phase, $\phi$, of the amplitude emanating from a single slit (top slit in Fig. \ref{fig:fig4}) \cite{Englert1996}. \\
\indent Consider the case of Fig. \ref{fig:fig1} when the transmission probability of the light incident  on the double slit, $t$, is equal to one.  The probability to detect at detector D1 ($D_1$) is equal to one and the probability to detect at D2 ($D_2$) is equal to zero. Thus, \textit{D} is equal to 1 and \textit{V} is equal to zero, which gives $D^2+V^2= 1$.  Afshar's idea is that a fine grid can be placed so that the distinguishiblity is not affected while the interference pattern must have a\textit{} non-zero fringe contrast.  To have a strong interference pattern, both slits should be balanced.  So let's consider $t=\frac{1}{2}$, then $V^2 = 1$ and $D^2 = 0$, E-G-Y holds, apparently this is not the case Afshar considered.  Consider an intermediate case, $t=\frac{1}{4}$, now $V^2 = 3/4 $, $D^2= 1/4$ again $D^2+V^2 = 1$.  More generally, the E-G-Y inequality ($D^2+V^2\leq 1$) holds for all values of \textit{t}.  Given that these ideal scenarios (point source slits and infinitesimal wires) do not violate E-G-Y, we will now consider finite size slits and wires. \\
\indent Our simulation was performed using Matlab and path integral techniques.  The details of the techniques are described in \cite{Pritchard,ericpaper} and the code and data are available upon request. The wave function, $\psi$, and kernel, $K$, have the following form: 

\begin{equation} \label{eq:3} \psi_f(x_f) = \int K(x_i,x_f)\psi_i(x_i)dx_i \end{equation} 
\begin{equation} \label{eq:4} K(x_i,x_f) = e^{\frac{2\pi i L_i}{\lambda}} \end{equation}
\begin{equation} \label{eq:5} L(x_i,x_f) = \sqrt{(x_f-x_i)^2+L_{1,2}^2} \enskip , \end{equation} 

\noindent where $x_{i,f}$ are position coordinates in one of the following planes: double slit, interference, or detector (Figs. \ref{fig:fig1} and \ref{fig:fig2}). $L_1$ and $L_2$ are the distances between the planes and $\lambda$ is the wavelength of the laser.  To calculate the \textit{V} and \textit{D}, one propagates the wave function in the double slit plane, $\psi_{ds}$, to the wave function in the interference plane, $\psi_{int}$, and then once more to the wave function in the detector plane, $\psi_{det}$. \\
\indent The transmission probability in the double slit plane, $t$, is varied for each slit as shown in Figs. \ref{fig:fig1} and \ref{fig:fig2}.  The left slit has transmission probability \textit{t} and the right slit has $1-$\textit{t} (\textit{t} taking values from 0 to 1).  The width \textit{d} of the wires  that are placed in the interference plane are also varied.  The affect of the wire grid is modeled by multiplying the wave function in the interference plane, $\psi_{int}(x)$, with a transmission function, $T(x)$.  The lens is described by multiplication with a quadratic phase $R(x)$: 

\begin{equation} \label{eq:6} T(x) = \sum_{j=0}^{2N} \Pi(x_j), \quad N \in \mathbb{N} \end{equation}  
\begin{equation} \label{eq:7} \Pi(x_j) = 1 - (H(x_j+d/2)-H(x_j-d/2))\end{equation}
\begin{equation} \label{eq:8} x_j = j(\frac{\lambda L_1}{s})-\frac{b}{2}, \quad j=0,1,2,..,2N \end{equation}     
\begin{equation} \label{eq:9} R(x) = e^{i\alpha x^2} \end{equation} 

\begin{figure}[H]
	\begin{center}
    	\textbf{Simulation of Distinguishiblity}\par\medskip
		\resizebox{0.8\hsize}{!}{\includegraphics*{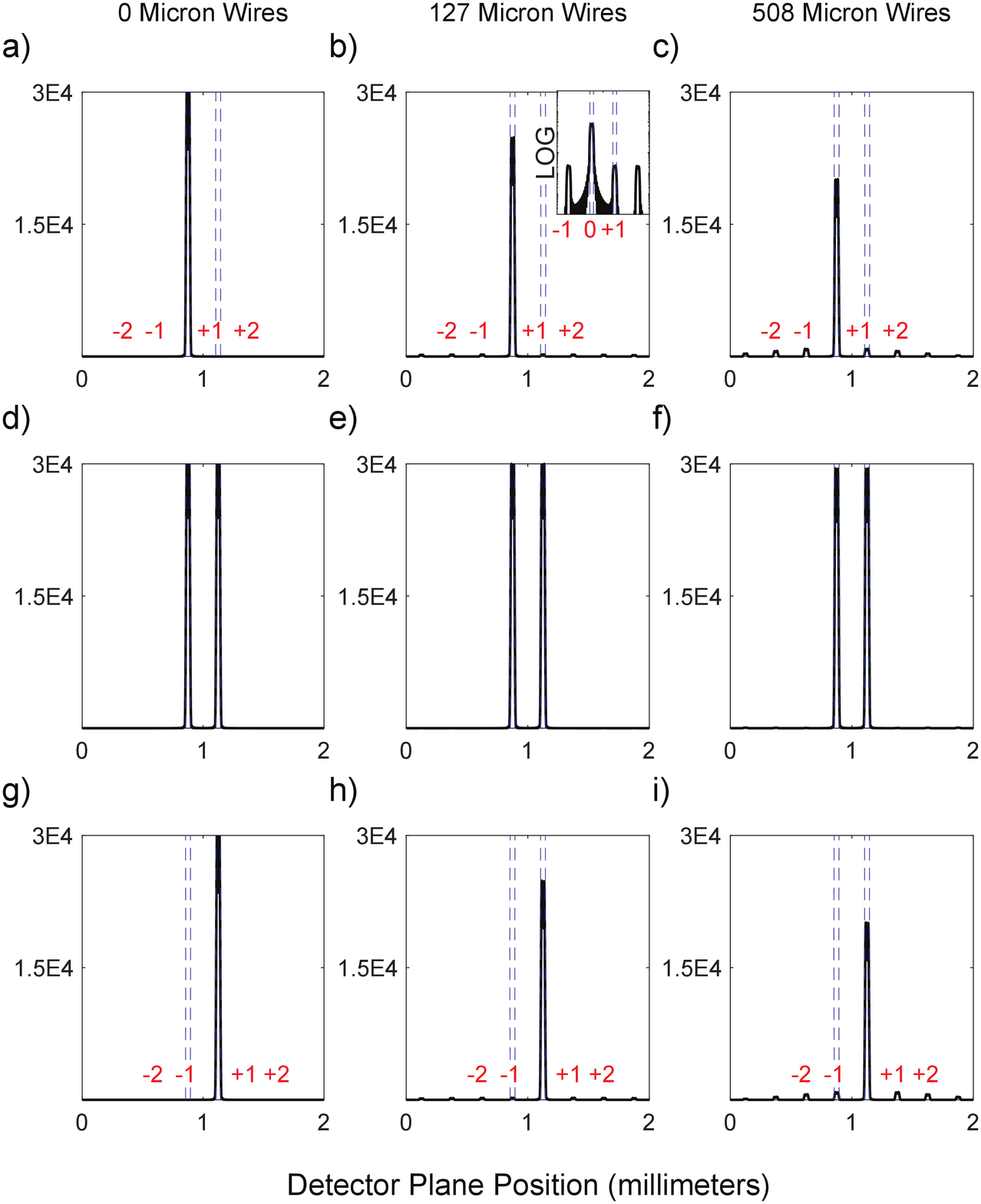}}
		\caption{Simulation of the Distinguishbility Measurement (setup of Fig. \ref{fig:fig1}).  The intensity is plotted on the vertical axis while the position in the detector plane is plotted on the x-axis.  From the left column to the right column, the bar thickness, \textit{d}, is varied.  From the top row to the bottom row, the transmission probability, \textit{t}, is varied from 0, 0.5, and 1.  The dashed blue lines represent the double slit projected onto the detector plane.  A log scale plot is added as an inset to 3b) in order to more clearly show the diffraction into the secondary peaks for the 127 micron wires.}   
	\label{fig:fig3}
	\end{center}
\end{figure}

\begin{figure}[H]
	\begin{center}
		\textbf{Simulation of Visibility}\par\medskip
    	\resizebox{1.0\hsize}{!}{\includegraphics*{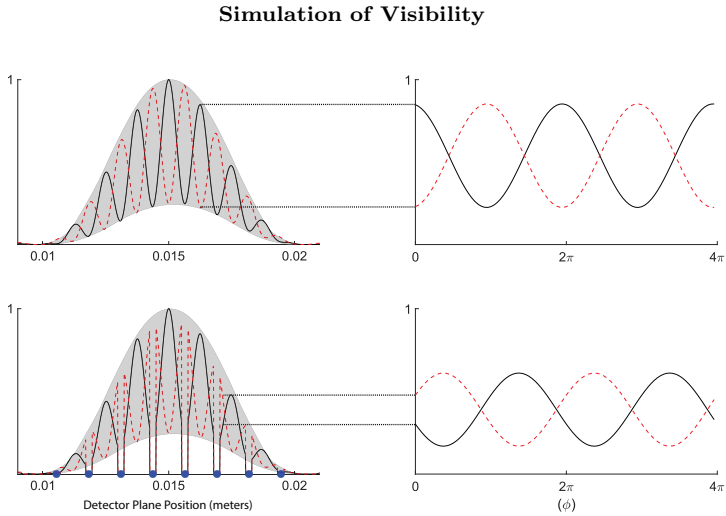}}
		\caption{Simulation of the Visibility Measurement (setup of Fig. \ref{fig:fig2}). The amplitude squared of the wave function, $|\psi(x)|^2$, is plotted vertically.  The x-axis of the left side panel represents the position in the detector plane, while the right side panel has an x-axis corresponding to the phase at a single location (indicated by a dashed line connecting the left and right panels).  The solid black line corresponds to a phase, $\phi=0$.  The dashed red line has phase, $\phi=\pi$. At any point, the phase can be smoothly varied (right side panels) to determine an envelope (grayed regions) that is indicative of the fringe contrast (or Visibility). The area directly behind the wires (blue circles) has no intensity and so the contrast cannot be measured there. Everywhere else, the contrast is the same, it is unaffected by the presence of the wire grid.}
	\label{fig:fig4}
	\end{center}
\end{figure}

\noindent The strength of the lens is determined by $\alpha$ and can be chosen such that the slits will be imaged in the detector plane. \textit{H} is the Heaviside function and $\Pi$ is a shifted Boxcar function. Interference minima occur in the interference plane at locations points $x_j$.  These points are in the far-field and are found in the usual way (eqn. (\ref{eq:8})), here \textit{b} is the width of the lens and \textit{s} is the slit separation. \\
\indent Afshar claims that: "no significant reduction in total radiant flux due to the wire is found"\cite{Afshar2007}.  He then claims that $D=1$ and that it is not affected by the thin wires.  At the same time, it is concluded that the insertion of the wires provides some knowledge on the presence of an interference pattern, and thus $V>1$. This would lead to $D^2+V^2 > 1$, a violation of the E-G-Y duality relation in one measurement.  The first problem is that two experiments are needed to measure \textit{D} and \textit{V} (Figs. \ref{fig:fig1} and \ref{fig:fig2}).  The second problem is that for both slits open, $D=0$ not 1.  Finally, \textit{D} is affected by the wires, the scattering can be seen in Fig. \ref{fig:fig3}.  Using the simulation we will show that there is a small amount of diffraction from the grid wires and that the inequality $D^2+V^2\leq 1$ is not violated.  This is consistent with the experiment of Jacques et al. \cite{jacques2008NJP}. \\
\indent To support the claim that \textit{D} is not affected by the grid, Afshar and Cramer point to diffraction data in the experiment \cite{Afshar2007,Cramer2016,floresresponse}.  The problem is that the diffraction peaks from the grid are very small for the parameters used and would be buried in the background noise.  To illustrate this, we compare the diffraction with three wire grid sizes. Fig. \ref{fig:fig3}a, d, g shows the diffraction image in the detector plane for $t=0, 0.5,$ and $1$, respectively, without the wire grid. The result of having a wire grid with 127 $\mu m$ bars is shown in Fig. \ref{fig:fig3}b, e, h.  The amount of diffraction into the smaller peaks of Fig. \ref{fig:fig3}b, e, h (labeled as -2, -1, +1, +2) is small compared to the main peak heights representing the image of the slits.  This means that the values of $D_1$ and $D_2$ in eqn. (\ref{eq:1}) have almost not changed. Examining Fig. 1 of Afshar's 2007 paper \cite{Afshar2007}, we estimate the noise in the experimental data to be 1\%. Comparing this to the height of the peaks in Fig. \ref{fig:fig3}b, e, h of the simulation, it is clear that the experimental data would not reveal the presence of diffraction.  When the wire grid size is increased by a factor of three (Fig. \ref{fig:fig3}c, f, i), the ratio of diffracted peak height (-2, -1, 1, 2) increases relative to the zeroth-order peak heights (small diffraction peaks can be seen in Fig. \ref{fig:fig3}f as well).

\begin{figure}[H]
	\begin{center}
		\textbf{Distinguishiblity vs. Visibility}\par\medskip
		\resizebox{0.9\hsize}{!}{\includegraphics*{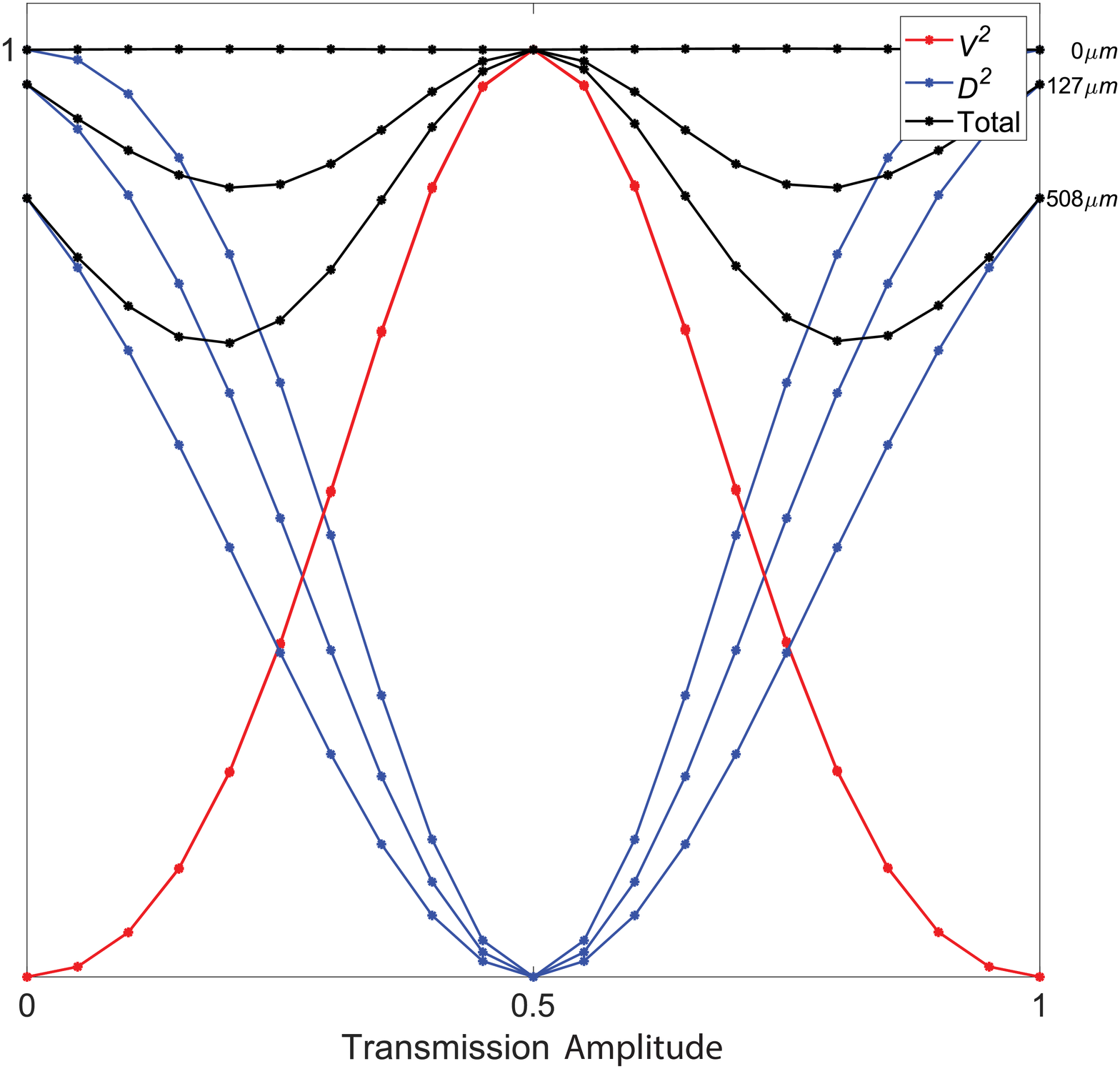}}
		\caption{Visibility and Distinguishably. The x-axis is the transmission amplitude for each slit (the far left, with an amplitude of zero gives a fully illuminated right slit and a blocked left slit).  Curves corresponding to different wire thicknesses are labeled on the right hand side.  For all cases $D^2+V^2\leq 1$.}
	\label{fig:fig5}
	\end{center}
\end{figure}

\indent Fig. \ref{fig:fig4} illustrates the affect that the wire grid has on \textit{V}. Previous estimates \cite{jacques2008NJP,steuernagel2007} found a \textit{V} that was altered upon insertion of the grid. Varying the phase between the slits as per the definition \cite{Englert1996,GreenbergYasin1988} shows no loss in fringe contrast.  This is illustrated by comparing the bounds of the grayed regions in the Fig. \ref{fig:fig4}.  The addition of the grid does not affect the envelope. In Jacques's experiment \cite{jacques2008NJP}, no phase shift was used, instead the wire grid was translated to estimate \textit{V}.  When using thicker wires, this estimate can vary significantly from the result using the formal definition. \\
\indent Afshar determines \textit{D} for the case when both slits are open (middle column of Fig. \ref{fig:fig3}) to be 1.  This is not consistent with the follow up analysis \cite{jacques2008NJP} or the original definition \cite{Englert1996,GreenbergYasin1988}.  In order to calculate \textit{D}, the setup in Fig. \ref{fig:fig1} is simulated and the results are determined using eqn. (\ref{eq:1}). For \textit{V}, a second simulation is needed (Fig. \ref{fig:fig2}) with eqn. (\ref{eq:2}) used to determine the results.  In both simulations, the transmission probability, \textit{t}, and bar width, \textit{d}, are varied.  The results are combined to create Fig. \ref{fig:fig5}. We find no violation of the E-G-Y duality relation $D^2+V^2\leq 1$ for any value of \textit{d} or \textit{t}. 

\section{Conclusion}
\label{sec:2}
\indent Our simulation shows that the Afshar experiment \cite{Afshar2007} does not constitute a violation of the E-G-Y \cite{Englert1996}\cite{GreenbergYasin1988} duality relation.  Consequently, claims that TIQM captures the physics of this experiment while the Copenhagen and Many-worlds interpretations do not \cite{Cramer2016}, should be reevaluated.

\section{Commentary}
\label{sec:3}

\indent We have reached out to John Cramer regarding his support of the Afshar experiment and specifically about the comments made in his 2016 book \cite{Cramer2016}. After sharing the results of our simulation, he responded by reiterating the comments made in his book: \\[0.2cm] "the amount of light intercepted by the wires is very small, consistent with 0\% interception. This implies that the interference minima are still locations of zero intensity and that the wave interference pattern is still present, even when which-way measurements are being made... This observation would seem to create problems for the complementarity assertions of the Copenhagen Interpretation" \cite{Cramer2016} \\[0.2cm] To address this statement we can interpret E-G-Y's wave-particle duality relation and Bohr's description of complementarity.  While using a lens, one measures the probability as a function of \textit{x}, which yields a value of \textit{D}.  Without a lens, one measures the probability as a function of $\phi$, which yields a value for \textit{V}.  The E-G-Y relation is satisfied in all cases considered (for all values of \textit{t}) We feel that the modern interpretation of Bohr's complementarity principle is captured in E-G-Y.  Cramer's discussion appears to focus on the Copenhagen interpretation as codified by Bohr.  We would like to point out that it seems Bohr intended his complementarity principle to only apply to the results of mutually exclusive experiments \cite{Cramer2016,Bohr}.  The measurement of \textit{D} and \textit{V} are indeed mutually exclusive in this example, as they cannot both be determined simultaneously using the same experimental arrangement. Thus the Afshar experiment doesn't violate Bohr's words as originally written: \\[0.2cm]
\indent "In Bohr’s words [43]: “ … we are presented with a choice of either tracing the path of the particle, or observing interference affects, … we have to do with a typical example of how the complementary phenomena appear under mutually exclusive experimental arrangements.”" (quote taken from John Cramer's 2016 book) \cite{Cramer2016} \cite{Bohr} \\[0.2cm]
\indent It is our opinion that Professor Cramer's Transactional Interpretation can help to develop an intuition for quantum processes and this work is not intended to diminish it's value.  Instead we hope that this work enhances the value of TIQM by countering such widely accessible (even if not rigorous) accounts as given in Wikipedia: "More recently, he has also argued TIQM to be consistent with the Afshar experiment, while claiming that the Copenhagen interpretation and the many-worlds interpretation are not." \cite{wiki} \\
\indent Students are often taught wave-particle duality by using example cases that result in an interference pattern or which-way information.  It is not typically discussed that there are cases when partial which way information is known and that in these cases, interference effects are still present.  The Afshar experiment provides a simple system in the context of diffraction that teachers could use to show that there is in fact a continuum of possibilities in agreement with the standard formalism of quantum mechanics. It could also help to elucidate the idea that interpretations of quantum mechanics are different from new theories in that they indicate what words scientists ascribe to the mathematics and are not discernible in experiment

\end{document}